\documentclass[prl, superscriptaddress, amssymb,nofootinbib,amsmath,showpacs,twocolumn, floatfix, balancelastpage]{revtex4-1}

\usepackage{graphicx}
\usepackage{psfrag}
\usepackage{enumerate}
\usepackage{amssymb}
\usepackage{amsfonts}
\usepackage{amsmath}
\usepackage{bold-extra}

\newcommand{\be}{\begin{equation}}
\newcommand{\ee}{\end{equation}}
\newcommand{\beq}{\begin{eqnarray}}
\newcommand{\eeq}{\end{eqnarray}}

\def\[{\left [}
\def\]{\right ]}
\def\({\left (}
\def\){\right )}

\def\r2{\sqrt{2}}

 \def\simleq{\; \raise0.3ex\hbox{$<$\kern-0.75em
      \raise-1.1ex\hbox{$\sim$}}\; }
   \def\simgeq{\; \raise0.3ex\hbox{$>$\kern-0.75em
      \raise-1.1ex\hbox{$\sim$}}\; }

\newcommand{\bbibitem}[1]{\bibitem{#1}\marginpar{#1}}

\newcommand{\figref}[1]{Fig. \ref{#1}}

\newcommand{\del}{\delta}

\newcommand{\gam}{\gamma}

\def\Label#1{\label{#1}%
  \smash{\hbox to0pt{\raise1ex\hbox{\tiny[#1]}\hss}}}
\def\noLabels{\let\Label=\label}
\def\nobbibitem{\let\bbibitem=\bibitem}

\begin{document}

\noLabels
\nobbibitem

\DeclareGraphicsExtensions{.pdf,.png,.gif,.jpg,.eps}

\title{Inflation from Flux Cascades}

\author{Guido D'Amico}

\author{Roberto Gobbetti}

\author{Matthew Kleban}

\author{Marjorie L Schillo}
\affiliation{Center for Cosmology and Particle Physics,  New York University, New York, NY 10003, USA}


\begin{abstract}
\noindent
%
When electric-type flux threads compact extra dimensions, a  quantum nucleation event can break a flux line and initiate a   cascade that unwinds many units of flux. Here, we present a novel mechanism for inflation based on this phenomenon.  From the 4D point of view, the cascade begins with the formation of a  bubble containing an open Friedmann-Robertson-Walker cosmology, but the vacuum energy inside the bubble is initially only slightly reduced, and subsequently decreases gradually throughout the cascade.  
If the initial flux number $Q_0 \simgeq {\mathcal O}(100)$, during the cascade the universe can undergo  $N \simgeq 60$ efolds of inflationary expansion with gradually decreasing Hubble constant,  producing a nearly scale-invariant spectrum of adiabatic density perturbations with amplitude and tilt consistent with observation, and a potentially observable level of non-Gaussianity and tensor modes. 
The power spectrum has a small oscillatory component that does not decay away during inflation, with a period set approximately by the light-crossing time of the compact dimension(s).  Since the ingredients are fluxes threading compact dimensions, this mechanism fits naturally into the string landscape, but does not appear to suffer from the eta problem or require fine-tuning (beyond the usual anthropic requirement of small vacuum energy after reheating).  

\end{abstract}






\maketitle

\noindent \textbf{\textsc{Introduction}}:

There is strong observational evidence for inflation.  However, understanding its microscopic origin requires knowledge of very high-scale physics.
Instead, most models of inflation are \emph{ad hoc} effective field theories, typically involving scalars with fine-tuned potentials.  
Attempts to realize slow-roll inflation in string theory are plagued by difficulties that require a complex set of ingredients to circumvent.  On the other hand, string theory naturally predicts the existence of a  very large number of eternally inflating meta-stable phases (the ``landscape''~\cite{Bousso:2000xa, Susskind:2003kw}) similar to those considered by Guth in his original ``old" model of inflation~\cite{Guth:1980zm}.   Such phases also occur  in many  theories besides string theory, and in all cases serve as  powerful attractors that dominate the system for essentially arbitrary choices of initial conditions.  These phases decay  via first-order phase transitions, but the bubbles produced by this process usually contain Friedmann-Robertson-Walker (FRW) cosmologies dominated by negative spatial curvature that do not reheat to a radiation dominated phase, and so old inflation was abandoned in favor of slowly rolling models~\cite{Linde:1981mu}.

One source for the vacuum energy driving false vacuum eternal inflation is higher form gauge flux (present in all string  and supergravity theories).   Here, we present a novel mechanism by which an eternally inflating false vacuum supported by flux can decay to a bubble containing a negatively curved FRW cosmology dominated by vacuum energy that subsequently steadily decreases over 60 or more Hubble times---much as in standard slow-roll inflation---before reheating via brane-anti-brane annihilation.  This post-bubble nucleation expansion inflates away the initially large negative curvature, and at the same time produces a scale-invariant spectrum of curvature perturbations that is consistent with observational data.  This occurs without the introduction of ad-hoc scalar potentials (indeed, without fundamental scalars at all), does not require any significant fine-tuning, and uses ingredients that are present in all string compactifications: higher-form flux, and at least one compact extra dimension. As such, this mechanism is both a natural candidate for the fundamental origin of slow roll inflation, and serves as a prototype for a new class of inflationary models.  
\paragraph{Relation to previous work:}  The idea of using a bubble collision in a compact dimension to reheat homogeneously and isotropically was first proposed by A.~Brown in~\cite{Brown:2008ea}, and the possibility of a scalar cascade in such a model was mentioned in \cite{Giblin:2010bd}. Discharge of higher-form flux by branes was first considered in  \cite{Brown:1988kg}, and flux cascades in \cite{Kleban:2011cs}.  Another model of inflation involving compact extra dimensions to extend the field range is \cite{Silverstein:2008sg}.

\smallskip

\noindent \textbf{\textsc{Mechanism}}:

The basic mechanism  is the ``flux discharge cascade" phenomenon recently discovered in~\cite{Kleban:2011cs}, where a $p+2$-form electric flux threading  at least one compact dimension can ``unwind'', repeatedly discharging in a cascade triggered by the quantum nucleation of a charged brane, and hence steadily decreasing the effective four dimensional vacuum energy.  The necessary ingredients are:
\begin{itemize}
\item A $(p+2)$-form field strength $F$ with $p \geq 3$, and $p$-branes  that are electrically charged under $F$.
\item A $D=4+q$ dimensional spacetime  $dS_{4} \times \mathcal M_{q}$, where $dS_4$ is 4D de Sitter spacetime and $\mathcal M_{q}$ is a stabilized compact $q$-manifold with  $q \geq 1$
\item $Q_0 \gg 1$ units of $F$ flux threading $dS_{4}$ and a $p-2$ cycle in $\mathcal M_q$, supplying the $dS_4$ vacuum energy 
\end{itemize}
The simplest realization---which for clarity we will focus on in this note---is a 5D space-time $dS_4 \times S_1$ with a bare cosmological constant, $Q_0$ units of initial 5-form flux, and some other source of energy (for instance Casimir energy of several bosonic and fermionic fields~\cite{ArkaniHamed:2007gg}) that stabilizes the $S_1$.  

\begin{figure}[t]
\hspace{-0.1 in}\includegraphics[width=0.32\textwidth]{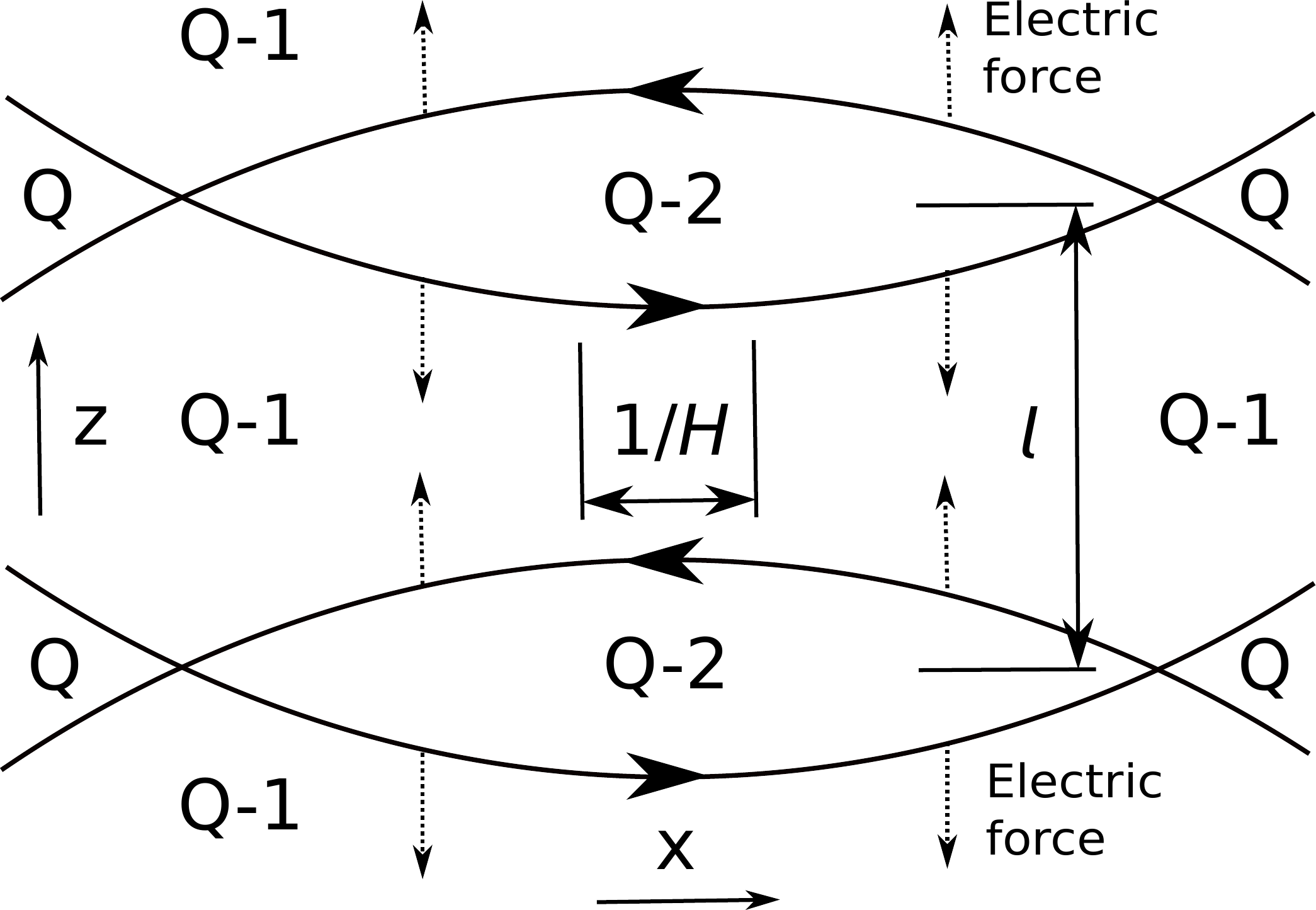}
\caption{\label{fig} A snapshot of the flux discharge cascade on a $dS_4 \times S_1$ space-time.   $Q, Q-1,...$ indicate the number of flux units, and the dashed arrows indicate the direction of the force (and velocity) of that section of brane.  Typically $l<1/H$; the figure is drawn not to scale for illustrative purposes.}
\end{figure}

Classically, a $D$-form flux in $D$ space-time dimensions (a ``top form") is simply a (positive) contribution to the cosmological constant.  However, $F$ can discharge locally by an analogue of the Schwinger process---the nucleation of an approximately spherical bubble of electrically charged brane \cite{Brown:1988kg}, which we assume to be smaller than the compact space $\mathcal M_q$, which itself we assume to be smaller than the Hubble length $1/H$ (since the contrary is difficult  or perhaps impossible to stabilize).  For a top form, the brane is co-dimension 1 in the space and the flux everywhere inside the brane bubble is constant, but reduced relative to the exterior by one unit of the brane's charge.  The electric flux at the bubble wall exerts a force on it, causing it to expand in all directions.  The bubble expands freely in the 3+1 dimensions of spacetime, but in the compact dimension(s) it wraps around and collides with itself.  Ignoring self-interactions for a moment, it passes through itself, and in the overlap region discharges the flux by two units.  This region expands and, after a second wrap, forms a region with three units discharged, etc. (\figref{fig}). 

An observer located at a point in the spacetime would encounter a series of brane walls that sweep across her location at regular intervals, and a flux that is constant except when a wall crosses her position, after which it decreases by one unit.  An observer unable to resolve distance or time scales of order the size of $\mathcal M_q$ would simply observe a steadily decreasing flux.  Because the flux contributes positively to the effective vacuum energy of the 4D spacetime, during the cascade there is a gradual decrease in the Hubble constant of the de Sitter---just as in ordinary slow-roll inflation.  After part or all of the flux is discharged, the remaining effective 4D vacuum energy can be positive, negative, or zero, depending  on the stabilization mechanism and any additional fluxes or vacuum energy.  For our purposes, we will assume the  4D vacuum energy at the end of the cascade is close to zero, or at the value that will result in nearly zero vacuum energy after GUT or standard model phase transitions that occur later in the evolution of the universe---a significant fine-tuning, but the only one needed, and the same one required for any successful model of inflation (or to solve the cosmological constant problem in the string landscape~\cite{Bousso:2000xa}).   

The reduction in flux during the cascade can also lead to a change in the geometry of $\mathcal M_q$, typically reducing its overall volume. However if the flux is not the primary element that stabilizes $\mathcal M_q$ this change is small, and we will neglect it.

\smallskip
\noindent \textbf{\textsc{Brane Nucleation and Expansion}}:

The discharge of $p$-form gauge fields by brane nucleation was first described in~\cite{Brown:1988kg}, and the cascade in~\cite{Kleban:2011cs}.  Our analysis is similar, with the  additional ingredients of a curved spacetime metric and extra compact dimension(s).  To study the decay, we should first find the instanton that corresponds to the nucleation of the bubble of brane.  We will focus on the simplest case  $dS_4 \times S_1$, with Euclidean signature metric $ds_E^2 = H^{-2}\left(d\xi^2 +\sin^2{\xi}d\Omega_{3}^{2}\right) + dz^2$, where $z \simeq z+l$.

Typically the dominant instanton for decay of a false vacuum state has the maximal symmetry possible.  We will assume the brane is thin, and that the instanton depends only on $\xi$ and $z$ in accord with the symmetry of the initial state.  We are interested in the case where the size $\Delta z$ of the instanton in the $z$ direction satisfies $\Delta z < l$, so  the periodic boundary conditions do not affect the solution (at least in the thin-wall limit).  Finally, when the initial number of flux units $Q_0 \gg 1$  the gravitational backreaction of a single bubble is small and can be ignored for the purpose of finding the instanton.

With these assumptions the instanton solution is fully characterized by the location of the wall $z=\pm z_b(\xi)$, where $\pm$ refers to two symmetric halves  and we have chosen $z=\xi=0$ as the center of the bubble.  To find $z_b(\xi)$, one should minimize the action $S_E = - \kappa\int_{V}\sqrt{g_E} + \sigma\int_{\partial V}\sqrt{g_{E(induced)}},$
where $ \kappa$ is the difference in energy density on the two sides of the wall, $\sigma$ is the tension of the wall, and $V$ is the volume enclosed by the bubble.

Extremizing this action results in equations of motion that can be solved analytically for $dz_b/d \xi$, with the integration constants fixed by the requirements of finite action and smoothness.  For brevity we will not reproduce the result here.  The coordinate shape of the instanton is oblate, in the sense that the maximum excursion in $z$ is less than that in $\xi$, and approaches spherical with radius $R_0=4 \sigma/ \kappa$ in the limit $R_0 \ll 1/H$.  

To find the time evolution of the bubble immediately after it nucleates, one  analytically continues the instanton solution using $\xi \rightarrow it$,  $d \Omega_3 \rightarrow i d H_3$, where $d H_3^2$ is the metric on a unit 3-hyperboloid.  After this continuation the metric is de Sitter spacetime in a hyperbolic slicing, times $S_1$ ($ds^2 = -dt^2 + \sinh^2 (t) d H_3^2 + dz^2$), while the analytically continued instanton---that now depends only on $t$---describes a spherical brane that expands in both the de Sitter and $z$ directions.  As in the case of Coleman-de Luccia tunneling, the interior of the bubble can be foliated by spacelike 3-slices in the de Sitter directions that are homogeneous and isotropic and have constant  curvature and energy density, and the resulting FRW cosmology is  dominated by negative spatial curvature immediately after the bubble forms.

There are  two crucial differences with standard Coleman-de Luccia tunneling.  The first is that the bubble expands in the extra, flat $z$ direction with a limiting asymptotic speed: $\lim_{t \rightarrow \infty}{dz_b}/{dt} \equiv v = {1}/{\sqrt{1+\(3\sigma H/ \kappa\)^2}}$~\cite{Brown:2008ea}.  In the models of interest $3 \sigma H/\kappa \ll 1$ and  $v$ is relativistic during the cascade.  The second difference is that because $z$ is compact,  the bubble will collide with an image bubble when $z_b=N l/2$ (for integer $N$).   For the same reason cosmic bubble collisions in 4D preserve an $SO(2,1)$ hyperbolic symmetry in the directions transverse to the collision axis (see \emph{e.g.} \cite{Kleban:2011pg}), these collisions occur on spacelike surfaces of constant $t$ (3-hyperboloids), taking place at approximately equally spaced instants during the cascade.  Crucially, \emph{ brane self-collisions do not disturb the homogeneity and isotropy of the open 4D FRW cosmology inside the bubble.}

After a few efolds of de Sitter expansion, the radius of curvature of the bubble (and equivalently, that of the hyperbolic constant-$t$ slices) becomes exponentially longer than the de Sitter length $1/H$ (\figref{fig}).  From this time forward the space-time metric can be well approximated by $ds^2 = -dt^2 + \exp(2 H t) d \vec x^2 + dz^2$, and the bubble by two disconnected parts: a planar (anti)brane extended in the $\vec x$ directions and located at $z=(-)+z_b(t)$, moving at close to the terminal velocity $(-)+v$.  Every interval of time $\Delta t = l/2v$, the brane and anti-brane collide.  

Such collisions can lead to a number of possible outcomes: perfect transmission or reflection, total annihilation, or transmission or reflection accompanied by some particle or string production.  In string theory at relativistic velocities the branes transmit, because the tachyonic mode of the open string (that represents the annihilation channel) does not have time to condense as they pass through one another \cite{us, Kleban:2011cs}.   There is some string production, which we will briefly mention later.

In general, assuming the branes transmit through one another, each collision produces a new region in between the planes where the flux has been reduced by one additional charge unit.  Averaged over timescales of order $l/v$, this corresponds to a decrease in the flux $F$ that is linear in time, or a decrease in the energy density $\rho \sim F^2$ that is quadratic in time---much like $m^2 \phi^2$ inflation.  Other dependences are possible with more complex compact geometries \cite{us}.

\smallskip
\noindent \textbf{\textsc{Effective Action}}:
In order to derive the Lorentzian equations of motion for the brane and its fluctuations, we will need a more explicit form of the action for the brane and flux.  As we will see, the position $z_b$ of the brane in the extra dimension plays the role of a 4D inflaton, with a Dirac-Born-Infeld  kinetic term~\cite{Alishahiha:2004eh} and an effective potential that is quadratic up to a small oscillating component.  The brane/flux action is
 \be \begin{split}
	S=& \int_{0}^{l} dz\int dt d^3 x ~ e^{3 Ht}\times  \\ & \left\{- 2 \sigma  \sqrt{1-(\partial_t z_b)^2+e^{-2 H t} (\partial_{\vec x} z_b)^2} \delta(z-z_b )
	 - \frac{F_5^2}{2 \cdot 5!} \right\},
\end{split} \ee
where coupling between the brane and the flux determines 
$$
\frac{F_5^2}{5!} =  {\mu^5} \left(Q_0 + \sum_{j=-\infty}^{\infty}\left[ \Theta(z-z_b+j l)-\Theta(z+z_b+j l) \right]\right)^2.
$$
Here $\mu^{5/2}$ is the charge of the brane, and the sum arises due to the periodicity of $z$.  The action can be integrated over $z$ to obtain a 4D effective action:
$$
S = \int dt d^3 x e^{3 H t} \( -2 \sigma \sqrt{1-(\partial_t z_b)^2+e^{-2 H t} (\partial_{\vec x} z_b)^2}  - V(z_b) \).
$$
Here $V(z_b)$ is a piecewise-linear approximation to a quadratic:
$$
\frac{dV}{dz_b} = - 2 \mu^5\left(Q_0 - \frac{1}{2} - \left[ \frac{2 z_b}{l} \right] \right),
$$
where $\left[ ... \right]$ denotes  integer part.

The position of the brane $z_b(t)$ can be found by solving the equations of motion that result from this action, with the result that $\dot z_b \approx {1}/{\sqrt{1+\(3\sigma H/ \kappa\)^2}}$ is nearly constant throughout the cascade.  This is because the change in energy density across the brane $ \kappa = \mu^5(Q-1/2)$, while $H \approx \sqrt{(8 \pi G_N/3) \rho_F} = \sqrt{(8 \pi G_N/3) \mu^5 Q^2/2}$ is close to linear in $Q$.  Hence $v$ is approximately constant and $z_b = \int v~ dt$ is linear in time, up to oscillatory corrections with period $l/(2 v)$ and amplitude suppressed by $1/Q$.

\smallskip
\noindent \textbf{\textsc{Perturbations}}:
In any model of inflation, one is interested in the curvature perturbations $\zeta$ on the reheating surface.  In our model, inflation proceeds until most or all of the initial flux $Q_0$ is discharged.  At that point the velocity $v$ of the brane and anti-brane becomes non-relativistic,  and the next collision will lead the branes to annihilate rather than transmitting each other, thereby reheating the universe and ending inflation.  The reheating surface is simply the position of the brane $z_b(\vec x)$ at the time this last collision occurs.  

If the total flux discharged is $Q_t$, reheating occurs when $z_{b}(t)+\delta z_b(\vec x, t)=v t + \delta z_b = Q_t l/2$.  Solving for $t$ yields $t=Q_t l/2 v-\delta z_{b}/v \equiv t_{0} + \delta t.$  The curvature of this hypersurface is $a(t_{0} + \delta t) = a(t_{0}) + \dot a(t_{0} )\delta t,$ and so
$\zeta = \delta a/a = H \delta t = H \delta z_b/v = H \delta z_b/\dot z_b$, where $H$ and $\dot z_{b}$ are evaluated at horizon crossing as usual.   Therefore, to find $\zeta$ one should compute the perturbations in the brane location $z_b$, for which there are several distinct sources.  

Since $ z_b$ is nearly massless,  de Sitter quantum fluctuations lead to perturbations in the usual way.  In addition, if the brane collisions are inelastic the energy going into particle production is a source of friction for the brane's velocity, and variations in the density of the produced particles will therefore source perturbations in $ z_b$ \cite{Green:2009ds, LopezNacir:2011kk}.

\paragraph{Scalar Perturbations:} To compute the amplitude and spectrum of the de Sitter perturbations we expand the action to quadratic order in the perturbation $\delta  z_b(t, \vec x)$ around the background solution $ z_b(t)$.  A straightforward calculation yields (dropping the subscript)
$$
\ddot{\del z} + 3 \( H + \gam^2 \dot{z} \ddot{z} \) \dot{\del z}
	- \frac{e^{-2 H t}}{\gam^2} \partial_{\vec x}^2 \del z = 0 ~ ,
	$$
where $1/\gamma^2 \equiv 1-\dot  z^2$.  This equation can be solved in momentum space giving Bessel functions; the result for late time is 
$
	\del z_k^2 =  {H^2}/({4 {\sigma k^{3}}})
$.
 Therefore the power spectrum of scalar curvature perturbations, ignoring any contribution from string or particle production, is
\be
\label{scalar}
	\mathcal{P}_{\zeta} = \frac{H^4}{8 \pi^2 \sigma v^2}.
\ee

\paragraph{Tensor Perturbations:}  The spectrum of tensor perturbations due to de Sitter fluctuations is simply 
\be
\label{tensor}
	\mathcal{P}_{h} = \frac{16 G_{N} H^2}{\pi}.
\ee
The tensor-to-scalar ratio is 
$$
r=\frac{128 \pi G_{N} \sigma v^{2}}{H^{2}}=  v^{2} \frac{R_0}{l} \frac{24}{Q},$$  
where again $R_0=4 \sigma/ \kappa$ is the approximate radius of the bubble when it formed. Since $R_0/l<1$ but not necessarily very small, and $Q \simgeq 10^2$, the tensor modes in this model are potentially observable in the near future.

\paragraph{Perturbations from String/Particle Production:}  To compute the contribution to $\delta z$ from variations in the number of strings or particles produced by brane collisions requires a model of the brane dynamics and degrees of freedom.  In string theory, stretched open strings will be pair-produced at each collision, because the mass of the stretched strings changes with time as the brane and anti-brane pass near or through each other. The rate can be calculated at one loop in $g_s$ by computing the imaginary part of the annulus diagram with appropriate boundary conditions.  Given the rate, perturbations in $\delta z$ can be computed using a technique similar to that of \cite{Green:2009ds, LopezNacir:2011kk}.  The result is that in a relatively broad range of string parameter space, particularly with $p=4$ or $5$ (4- or 5-branes), string production is subdominant to the largely model-independent de Sitter perturbations  \eqref{scalar} \cite{us}. With other choices of parameters string production is dominant, but in the interest of brevity we will not describe the results here.  

Brane collisions may also create tensor perturbations, either directly from Bremsstrahlung or from the decay of produced particles or strings, and the amplitude of these perturbations might exceed the de Sitter contribution \eqref{tensor} \cite{Senatore:2011sp}.  It would be interesting to investigate this possibility further.

\paragraph{Tilt:}  Because the variation in $v$ during the cascade is very small, the tilt of the scalar spectrum $n_s-1=d \ln P/d ln k$  arises almost entirely from the $H^4$ term in \eqref{scalar}:
\be \label{tilt}
n_s-1 \approx 4 \frac{\dot H}{H} \frac{dt}{d \ln k} \approx 4 \frac{\dot H}{H^2} \approx -2/N_*,
\ee
where $N_* = \int H dt = \int dH (H/{\dot H} ) \approx -H^2/(2 \dot H) $ is the number of efolds from the time the quadrupole mode crossed the horizon during inflation to the end of inflation.  In a model with a high reheat temperature $N_* \sim 60$, and  the tilt \eqref{tilt} is then  consistent with observation.

\paragraph{Non-Gaussianity:}  The largest non-linearity in the effective action is the Dirac-Born-Infeld kinetic term~\cite{Alishahiha:2004eh,Senatore:2009gt} which produces non-Gaussianity of equilateral type, with an $f_{\rm NL} \sim (1-c_s^2)/c_s^2$.  Higher dimensional generalizations of this model may also produce local non-Gaussianity at reheating, as we will discuss below.

\paragraph{Reheating:}  When most or all of the flux has been discharged, the Lorentz force that keeps the branes moving around the circle is greatly reduced.  At this point the various dissipative and attractive forces on the brane and anti-brane (for instance, those from stretched strings) will bring them to close proximity at low velocity, at which point they will annihilate and convert an energy density $2 \sigma$ into particles.  While the details of this annihilation are non-perturbative, the dynamics must be local on Hubble scales, and so long as it does not affect super horizon perturbations is largely irrelevant to the cosmological predictions of the model.  Rare regions in which different amounts of flux have discharged may arise, and these regions may expand, collide, and generate perturbations on larger than Hubble scales (possibly producing gravitational waves, as in \cite{Kosowsky:1992rz}).  While the probability of such regions seems exponentially small (because $\delta  z_b \ll l/2$), the dynamics of this is complex and remains to be carefully investigated.

\smallskip
\noindent \textbf{\textsc{Generalizations}}: The model we focused on here is the simplest possible, with one extra dimension and a $p=5$-form field strength.  However, the phenomenon of flux cascades extends to a  broad array of models with different values of $p$ and numbers of compact dimensions~\cite{Kleban:2011cs}.  

If the brane has co-dimension greater than 1 (\emph{i.e.} if $p<D+2$), there may be additional light scalars that describe the position of the brane in the transverse dimensions.  Along with $ z_b$---the radius of the brane bubble in the compact dimension(s) the flux threads---these additional scalars determine the time of re-heating (because to annihilate, the brane and anti-brane must come  close together).  Therefore, fluctuations in these scalars convert into curvature perturbations in a manner similar to that of~\cite{Dvali:2003em, Senatore:2010wk}, and could produce a potentially significant level of local non-Gaussianity.  

Another difference arises when $p>3$, so that the compact cycle the flux threads is more than 1D.  In such cases the geometry of the cycle is important.  For instance, with $p=4$ the cycle can be an $S_2$, in which case the brane is a string on the $S_2$ that repeatedly oscillates  from pole to pole,  wiping away one unit of flux on each pass \cite{Kleban:2011cs}.  In this case the self-collisions are qualitatively different, because at least classically the brane  shrinks to a point and  inverts.  Furthermore, the curvature of the $S_{2}$ affects the dynamics of both the brane and its perturbations.  Further investigation of the higher-dimensional versions of this model may prove very interesting.  

\smallskip
\noindent \textbf{\textsc{Conclusions}}:

We have presented a novel class of microscopic inflationary models that incorporates many effective models of inflation previously considered, including DBI inflation \cite{Alishahiha:2004eh}, trapped \cite{Green:2009ds} and dissipative \cite{LopezNacir:2011kk} inflation, and hybrid inflation.  It provides a building block for new effective theories, and can potentially be naturally embedded in string theory.  Our inflationary mechanism does not require significant fine-tuning of either the parameters or the initial conditions, and naturally produces a spectrum of perturbations that is consistent with observation.  All models in this class predict an interesting and characteristic set of features in the power spectrum, including  oscillatory components and potentially observable non-Gaussianities and tensor modes.  

\smallskip
\noindent \textbf{Acknowledgements}:
MK is supported by NSF CAREER grant PHY-0645435 and NSF grant PHY-1214302.
G. D'A. is supported by a James Arthur Fellowship. The authors thank A.~Brown, S.~Dubovsky, G.~Dvali, R.~Flauger, B.~Freivogel, V.~Gorbenko, A.~Hebecker, A.~Lawrence, J.~Maldacena, L.~McAllister, M.~Roberts, L.~Senatore, T.~Tanaka, T.~Weigand, and M.~Zaldarriaga for useful discussions. 

\bibstyle{apsrev4-1}

\bibliography{Bibliography}

\end{document}